\documentstyle[12pt,epsfig]{article} 
\newlength{\dinwidth}
\newlength{\dinmargin}
\newcommand{\mr}[1]{{\mathrm {#1}}}

\newcommand{\reff}[1]{(\ref{#1})}
\newcommand{\be}{\begin{equation}}
\newcommand{\ee}{\end{equation}}
\newcommand{\bear}{\begin{eqnarray}}
\newcommand{\eear}{\end{eqnarray}}
\newcommand{\bc}{\begin{center}}
\newcommand{\ec}{\end{center}}
\newcommand{\bd}{\begin{description}}
\newcommand{\ed}{\end{description}}
\newcommand{\bit}{\begin{itemize}}
\newcommand{\eit}{\end{itemize}}
\newcommand{\ben}{\begin{enumerate}}
\newcommand{\een}{\end{enumerate}}

\def\lsim{\mathrel{\rlap{\lower4pt\hbox{\hskip1pt$\sim$}}
    \raise1pt\hbox{$<$}}}                
\def\gsim{\mathrel{\rlap{\lower4pt\hbox{\hskip1pt$\sim$}}
    \raise1pt\hbox{$>$}}}                
\begin{document}
\vspace*{1cm}
\begin{center}  \begin{Large} \begin{bf}
Virtual Photon Structure from Jet Production at HERA
\\
  \end{bf}  \end{Large}
  \vspace*{5mm}
  \begin{large}
J. Ch\'{y}la$^{a*}$, J. Cvach$^{a*}$\\
  \end{large}
\end{center}
$^a$ Institute of Physics, Na Slovance 2, Prague 8,
     18040, Czech Republic\\
$^*$ Supported by grants No. A1010619 and A1010602
of the GA of the Academy of Sciences of CR\\
\begin{quotation}
\noindent
{\bf Abstract:}
The feasibility of measuring parton distribution functions of
of virtual photons via the jet production at HERA is investigated.
\end{quotation}
Production of jets in ep collisions at HERA offers one of the ways of
studying the structure of the virtual photon. Due to the fact that this
structure needs time to develop,
parton distribution functions (PDF) $f_i(x,P^2,Q^2)$ of the virtual
photon, probed at the hard scattering scale $Q$, are expected to be
decreasing functions of the magnitude $P^2$ of its virtuality. As
$P^2$ approaches $Q^2$ from below $f_q(x,P^2,Q^2)$ should
approach parton model formula
\be
f_i^{\mr{PM}}(x,P^2,Q^2)\equiv
\frac{3\alpha}{2\pi}\left[x^2+(1-x)^2\right] \ln\frac{Q^2}{P^2}.
\label{PM}
\ee
The transition of the quark distribution functions of the real
photon to the form \reff{PM} is so far not calculable but
there are models which interpolate between $f_q^{\mr{real}}$ and
\reff{PM}.
Measurement of the PDF of the virtual photon would provide valuable
new information on the properties of parton interactions at
short distances. In ref. \cite{GD} this transition is parametrized as 
follows
\be
f_q(x,Q^2,P^2)=f_q^{\mr{real}}(x,Q^2)L(Q^2,P^2,\omega),\;\;\;
L\equiv \frac{\left[\ln(Q^2+\omega^2)/(P^2+\omega^2)\right]}
             {\left[\ln(Q^2+\omega^2)/(\omega^2)\right]},
\label{suppression}
\ee
while in analogous relation for the gluon $L$ is replaced by $L^2$.
The parameter $\omega$ determines the value of $P^2$ above which
the suppression factor $L$ becomes significant. Small $\omega$
means strong suppression already for weekly off--mass shell photons,
while $\omega\rightarrow\infty$ corresponds to no suppression at all.
The suppression formula \reff{suppression} is implemented, for instance,
in the recent versions of HERWIG MC generator.
Because of a different $x$ behaviour of PDF of the real photon and 
parton model formula \reff{PM}, $\omega$ must in general be a function 
of $x$.  As, however, in our simulations 
$P^2\ll Q^2\approx 4(p_T^{\mr{jet}})^2\ge100$ GeV$^2$ 
we considered $\omega$ as $x$ independent.

Jet production at HERA for general values of $P^2$ and $p_T^{\mr{jet}}$ 
is a two--scale problem, where it is thus not obvious what 
to take for the relevant har--scattering scale $Q^2$ in 
$f_i(x,P^2,Q^2)$: $P^2$ or $p_T^2$ of the produced jets, 
 or some combinations thereof? In this study we stayed in the 
region $\Lambda_{\mr{QCD}}\ll P^2\ll p_T^2$ and therefore assumed
$Q^2=\kappa p_T^2$ with the proportionality factor of the order of unity.  
To make the experimental procedure of jet finding well--defined  
and ensure the applicability of perturbative QCD, we furthemore required
$p_T^{\mr{jet}}\ge p_T^{\mr{min}}=5$ GeV, $P^2\le 10$ GeV$^2$.
As in this region dynamics of the jet production 
is close to that of the real photoproduction, all our further 
considerations were carried out in the $\gamma^{*}$p CMS. 
To study the $P^2$ dependence we 
split the region $P^2\le 10$ GeV$^2$ into eight intervals:
$(0,0.1)$ (``photoproduction''),
$(0.1,0.2)$, $(0.2,0.5)$, $(0.5,1)$, $(1,2)$, $(2,5)$, $(5,10)$
GeV$^2$, each with roughly the same number of events.
To guarantee good electron identification, the cut on 
$0.2\le y\le 0.7$ was imposed in all simulations.

Our simulations were guided by recent preliminary H1 and ZEUS data on
virtual photon structure \cite{H1,ZEUS}. The results presented here
are based on HERWIG 5.8d MC generator and standard
cone jet finder with $R=1$. 
To estimate the dependence of the results on the strength of the 
virtual photon suppression factor
the simulations were performed for three values of $\omega=0.1,1.0,3.0$.
We addressed the following questions: \\
a) How to isolate the
contribution of the resolved photon, which depend on $f_i(x,P^2,Q^2)$
and \hspace*{0.5cm} 
thus in principle allow its measurement, from direct photon 
one?\\
b) What are the ensuing requirements on the detector and
experimental procedure? \\
c) What is the required luminosity upgrade to get a
reasonable statistics? 

Most of the current attempts at separating resolved from direct photon
contributions to jet cross--sections are based on the fact
that for the latter the distribution of the variable
\be
x_{\gamma}\equiv
\frac{E_T^{(1)}\exp(-\eta^{(1)})+E_T^{(2)}\exp(-\eta^{(2)})}
{2E_{\gamma}},
\label{xgamma}
\ee
where $\eta^{(j)}$ and $E_T^{(j)}$ correspond to two jets with highest
transverse energies,
peaks at a value close to unity, while for the resolved component
the spectrum peaks at low $x$ and
drops rapidly as $x_{\gamma}\rightarrow 1$.
In parton model $x_{\gamma}$ is interpreted as a fraction of the
photon momentum carried by the parton or photon
participating in the hard collision with a parton from the proton.
In the direct channel and for two final state massless partons
$x_{\gamma}=1$ identically. Taking into
account nonzero $P^2$ leads to slightly modified formula for $x_\gamma$
but we sticked to \reff{xgamma} as in realistic QCD--based MC
simulations there are other, more 
important, effects that lead to the smearing
of the $x_{\gamma}$ distribution. To see which of them is most
important we compared, in both direct and resolved channels,
our MC results for
a) two final massless partons with no parton showers,
b) two final partons after they acquire nonzero virtuality,
c) jets formed out of final state on mass--shell partons and finally
d) realistic hadron jets. It turns out that the most
dramatic effect of the smearing, due mainly to hadronization,
occurs for the $x_{\gamma}$ distribution: instead of a pronounced
peak for $x_{\gamma}=1$ we get much wider and less pronounced structure
peaked at about $x_{\gamma}=0.85$, as shown in Fig.1a. Its
position and shape is essentially independent of $P^2$.
\begin{figure}
\epsfig{file=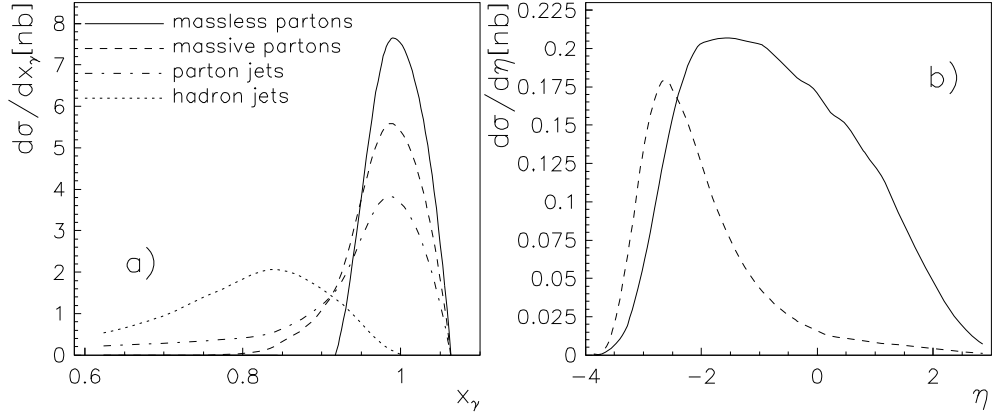,width=15cm}
\caption{a) Distributions of $x_{\gamma}$ in direct channel for $2<P^2<5$
GeV$^2$, taking into
account various smearing effects. b) The $\eta$ distributions
of jets with $E_T>5$ GeV in direct (dashed curve) and resolved 
(solid curve) channels.}
\end{figure}

\begin{figure}[bht]
\begin{center}
\epsfig{file=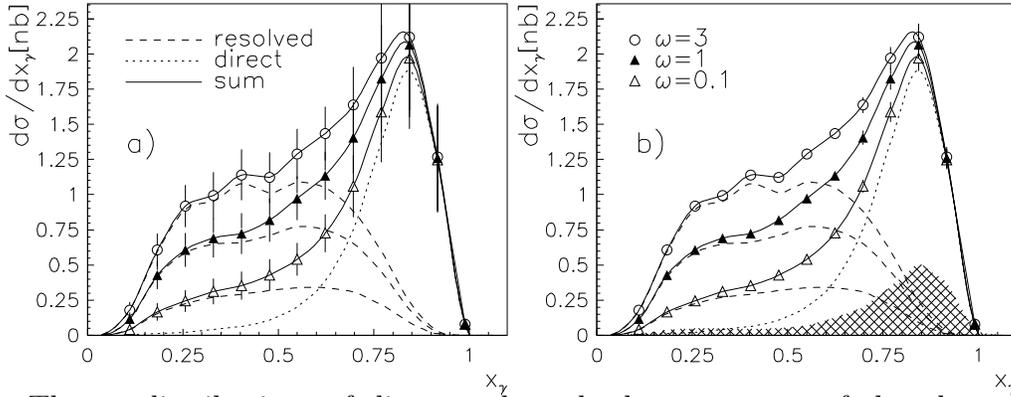,width=15cm}
\end{center}
\caption{The $x_{\gamma}$ distributions of direct and resolved
components of the photon and their sum for three values of
$\omega$ and jets with $\eta<-0.5$.
Superimposed are present (a) as well as anticipated future (b)
statistical error bars. The hatched in b) area shows 
systematic error due to 3\%
uncertainty in jet energy measurement.}
\end{figure}
To measure the parton
structure of the virtual photon requires a suitable
signature to separate resolved and direct components. The best
candidate remains, even after the smearing shown in Fig. 1a, the
$x_{\gamma}$ distribution. The resolved component can be enhanced by
imposing cuts on other variables. The most effective would be
a cut on the pseudorapidity $\eta>0$, illustrated in Fig. 1b.
Unfortunately, in this region there are problems with the separation of 
hard jets from the proton remnant one. Both 
experiments \cite{H1,ZEUS} therefore restrict their jets to the region 
$\eta< -0.5$. Another, but less effective way of enhancing the 
resolved component exploits the fact that the $\mid\!\Delta\eta\!\mid$
distribution is broader for the resolved component.  In some simulations
we therefore imposed also the cut $\mid \Delta\eta\mid>1$.

To assess the feasibility of measuring PDF of the virtual photon 
at HERA and
to get some idea of what the theoretical predictions look like, we show
in Fig.2 for three values of the suppression parameter
$\omega=0.1,1,3$ our MC results for the $x_{\gamma}$ distribution.
We see that the direct component
of the virtual photon gives rise to a peak at about
$x_{\gamma}=0.85$, while the resolved one, wherefrom the virtual photon
PDF would be determined, is dominant below $x_{\gamma}\approx
0.5$. The cross--over point, where the two contributions are equal
depends on $\omega$ and $P^2$ but lies around
$x_{\gamma}^{cr}=0.75$. The peak of the direct photon
contribution at $x_{\gamma}=0.85$ is reflected in the $\eta$
distribution (not shown) as the dominance of the direct component in
the low $\eta$ region around $\eta\approx -3$. 
The error bars superimposed in Fig. 
2a on the MC results characterize the present statistical errors, while
those in Fig. 2b indicate the effect of increasing the present 
luminosity by a factor of 50 to $50$ pb$^{-1}$. This increase would 
allow rather detailed study of $P^2$ dependence of overall suppression 
factor.  To measure the $x_{\gamma}$ dependence of the virtual 
photon PDF would, however, require still significantly higher 
luminosity.

\begin{figure}[th]
\begin{center}
\epsfig{file=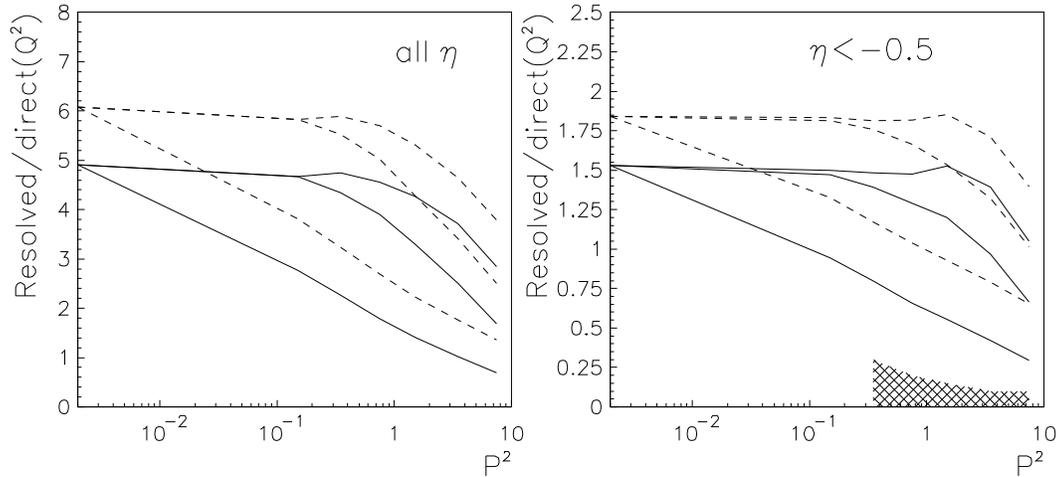,width=15cm}
\end{center}
\caption{$P^2$ dependence of the ratio
$R\equiv \sigma^{\mr{resolved}}/\sigma^{\mr{direct}}$.
The hatched area shows the systematic error due to 3\%
uncertainty in jet energy measurement. Solid lines correspond to $R$
as given by generator, the dashed ones to the method
based on the $x_{\mr{cr}}$ cut--off described in the text.
The triplets of solid and
dashed curves correspond from above to $\omega=3,1,0.1$.}
\end{figure}

The crudest measure of the resolved photon contribution to jet
cross--sections is the ratio
$R\equiv \sigma^{\mr{resolved}}/\sigma^{\mr{direct}}$.  
It depends, beside $\omega$,
sensitively on $p_T^{\mr{min}}$ and  also on 
cuts on $\eta$. We consider $p_T^{\mr{min}}=5$ GeV as is the minimal 
reasonable lower cut--off on $p_T^{\mr{jet}}$. Increasing 
$p_T^{\mr{min}}$ would significantly improve the possibility of 
separating direct and resolved components but, on the other hand, 
lower the statistics. In Fig. 3 we plot the ratio $R$ as a 
function of $P^2$ for three values of $\omega$. The solid curves 
correspond to $R$ evaluated from the knowledge, available in MC
generators, of 
separate contributions of direct and resolved channels. In real
experiments at HERA we may 
attempt to separate them using the cut on $x_{\gamma}$, defining the 
resolved contribution by the condition
$x_{\gamma}\le x_{\gamma}^{\mr{crit}}=0.75$
and complementarily for the direct one. The corresponding
results for $R$ are shown as dashed curves in Fig. 3.

\vspace*{0.2cm}
\noindent
{\large \bf Conclusions:}
\bit
\item
Higher luminosity is clearly a precondition to serious studies of
virtual photon structure.
\item Kinematical region of  positive $\eta$ in
hadronic CMS and large $\mid\Delta\eta\mid$ can further enhance the
contribution of the resolved component.
\item Direct photon component should be observable at about
$x_{\gamma}\doteq 0.85$.
\item Generator dependence should be investigated.
\eit

\end{document}